\documentclass[conference, 10pt]{IEEEtran}
\IEEEoverridecommandlockouts

\def\BibTeX{{\rm B\kern-.05em{\sc i\kern-.025em b}\kern-.08em
    T\kern-.1667em\lower.7ex\hbox{E}\kern-.125emX}}

\usepackage{cite} 
\usepackage{amsmath,amssymb,amsfonts} 
\usepackage{algorithmic} 
\usepackage{graphicx} 
\usepackage{textcomp} 
\usepackage{xcolor} 
\usepackage{comment} 
\usepackage{footnote} 
\usepackage[hidelinks,bookmarks=false]{hyperref}
\usepackage{gensymb} 
\usepackage{multirow} 
\usepackage{fancyhdr} 
\usepackage{subcaption} 
\usepackage{gensymb} 
\usepackage{tabularx} 

\usepackage{booktabs}
\usepackage{tabularx}
\usepackage{array}
\usepackage[none]{hyphenat}
\usepackage{siunitx}
\usepackage{multirow}
\usepackage{caption}
\usepackage{subcaption}  
\usepackage{makecell}  
    
\begin{document}

\title{Case for Vehicle-Edge Collaborative 
 Multi-Sensor Data Fusion for Autonomous Vehicle Teleoperation%
\thanks{ISBN 978-3-903176-82-9 © 2026 IFIP}}

\author{
\IEEEauthorblockN{Qixin Zhang, Ajay Kumar Gurumadaiah, Wei Ye, Eman Ramadan, Zhi-Li Zhang}
\IEEEauthorblockA{University of Minnesota -- Twin Cities, Minneapolis, USA \\
\{zhan8548, gurum021, ye000094, eman, zhzhang\}@umn.edu}
}

\maketitle

\begin{abstract}

Teleoperation provides a critical safety fallback when autonomous vehicles (AVs) encounter scenarios that are outside their operational design domain.
In practice, however, remote operators rely primarily on compressed camera streams over 5G, which often lack depth and spatial geometric cues for safe operation in complex dynamic environments. While multi-sensor fusion can enhance situational awareness, directly transmitting raw camera and LiDAR data is impractical due to 5G uplink bandwidth and latency constraints.
In this paper, we propose \textit{SHARDED}, a collaborative camera–LiDAR perception framework that deploys a feature-level fusion pipeline across the vehicle and edge to reduce uplink traffic while preserving 3D detection and depth estimation accuracy. 
We further design two complementary mechanisms for SHARDED: (i) A network-aware adaptive feature transmission mechanism that reduces data traffic by 50\% on average (peaking at over 95\%) compared to raw sensor data, and (ii) a latency-aware positional drift compensation mechanism to mitigate cross-modal misalignment induced by unstable network conditions. 
Evaluations on the nuScenes dataset and real-world 5G measurement traces show that SHARDED achieves competitive perception quality while reducing uplink bandwidth consumption and end-to-end latency.

\end{abstract}

\section{Introduction}
Autonomous vehicles (AVs), including robotaxis, are increasingly being deployed in urban environments. At Society of Automotive Engineers (SAE) Level~4 or below~\cite{2016-01-0128,sae2018taxonomy}, current AV systems operate within predefined operational design domains (ODDs), relying on accurate perception and prior knowledge (e.g., HD maps and route planning) for safe operation. However, unexpected and dynamic road conditions, including traffic accidents and road construction, can easily violate these assumptions and prevent reliable autonomous operation~\cite{Tangermann2025Waymo-incidents,WaymoAirportCrash2025, TeslaAustinCrash2025}. To handle such situations, teleoperated driving (TOD), in which a human operator remotely controls the vehicle, has emerged as an important fallback mechanism~\cite{remote-driving-of-autonomous-vehicles-Challenges-and-solutions,how-self-driving-cars-get-help,guident}. Providing \emph{real-time situational awareness}~\cite{camera-situation-awareness-1,camera-situation-awareness-2} 
is critical to AV teleoperation; this mandates streaming of sensor data, such as camera feeds, to a remote human operator.

TOD places stringent demands on cellular networks for remote perception, as operators must rely on transmitted perceptual data to make timely and safe control decisions.
In today's 5G networks, while the downlink throughput can achieve several Gbps, the uplink throughput is considerably lower, often around 100 Mbps or less, even under good channel conditions~\cite{Depth-Study-5G-mmWave-Networks,Ross-SIGCOMM24,We-4G-5G-Comparison-journal-paper}. This constraint limits both the volume and the variety of sensor data that can be effectively streamed over 5G networks. Furthermore, the sensor data must be delivered in a timely fashion, within $\leq$~100~ms per the 5GAA TOD requirement analysis~\cite{5GAA-ToD-system-requirements-analysis}.

Under current 5G capabilities, streaming 2D video from one or more cameras is the only practical option for AV teleoperation\footnote{The ``raw'' data rates generated by LIDAR are typically 100s Mbps. 
Even with existing state-of-the-art LIDAR compression schemes,  streaming LiDAR streams over commercial 5G networks is largely infeasible~\cite{zhang2022uplink,VENTIS23}.}. However, 2D video lacks depth information, making it difficult for a human teleoperator to accurately estimate object distances. This increases cognitive load and complicates safe vehicle control -- particularly in dense or dynamic traffic environments. These challenges are further exacerbated by fluctuations in video quality caused by bitrate adaptation mechanisms that cope with variable 5G bandwidth conditions. In addition, video compression artifacts can degrade the performance of object detection and tracking algorithms, which are often deployed at the teleoperation station to alert operators to emerging objects of interest~\cite{Qixin-CQR}. Such algorithms are critical for maintaining safe and reliable remote driving.

One way to combat these challenges is to augment 2D video with depth information extracted from LiDAR data, or, more generally, to augment 2D video with 3D object detection and tracking by performing sensor data fusion onboard the vehicle. The augmented 2D video (with depth or detected/tracked object data) is then delivered to the teleoperation station. However, as data fusion can take considerable processing time (in 10s ms up to over 100 ms); this plus the time to deliver the detection results over a 5G network mean that it may be too late to alert the teleoperator of an impending object. A perhaps more promising approach is to split a data fusion model\footnote{With neural network-based AI models, in theory, one can split a model along any layer of the model architecture.} across the vehicle and the edge cloud (where the teleoperation station resides), with portions of the model running on each side. Where and how to split a data fusion model has significant implications for compute and network bandwidth requirements, as well as overall latency performance. In other words, the efficacy of this approach hinges critically on available compute resources, the latency overheads of the \emph{split} fusion models, the fluctuating conditions of 5G networks, and the amount of data (e.g., intermediate features) that must be delivered over them.

In this paper, we explore multi-sensor data fusion for AV teleoperation.   In computer vision and AI fields, 
various multi-sensor fusion  models (especially for camera and LiDAR data fusion)
such as BEVFusion~\cite{liu2022bevfusion}, CMT~\cite{chen2022cmt}, and CLOCs~\cite{clocs}  
have been developed using three fusion approaches: early fusion, intermediate (or feature-based) fusion, and late fusion (see Section~\ref{sec:why_fusion}).  These approaches differ in when data (e.g., raw input, intermediate features, or results) extracted from each sensing modality is integrated or fused to accomplish the perception task (e.g., 3D object detection). Existing models are designed primarily for executing on a single machine (e.g., onboard a vehicle), regardless of the fusion approach used.  Leveraging these advances and recognizing these models can be deployed either \emph{all-on-vehicle}, \emph{all-on-edge}, or \emph{split across the vehicle-\&-edge}, we investigate multi-sensor fusion \emph{deployment} strategies for TOD perception pipeline design by systematically considering the key factors and design tradeoffs involved, such as onboard compute resources, (split) model overheads,  network bandwidth requirement, and overall TOD latency performance. Prior vehicle--edge collaborative perception and split inference approaches mainly target machine-only autonomy and typically assume fixed partitioning or offloading. In contrast, SHARDED is designed for human-in-the-loop TOD under practical 5G dynamics, explicitly addressing uplink-constrained feature transmission and latency-induced cross-modal misalignment that are not considered in prior approaches.

\textbf{Main Contributions}:
(1) We advanced SHARDED, a collaborative camera–LiDAR perception framework for teleoperated driving. SHARDED employs feature-level perception across the vehicle and edge to jointly optimize transmission, fusion placement, and timeliness under uplink constraints, while explicitly addressing uplink-constrained feature transmission and latency-induced cross-modal misalignment under dynamic 5G conditions.

(2) We designed a network-aware adaptive feature transmission mechanism with distance-based utility instantiation that prioritizes task-relevant spatial regions and significantly reduces uplink traffic, achieving up to 95\% data reduction compared with raw sensor transmission while preserving teleoperation-critical perception quality.

(3) We introduced a latency-aware positional drift compensation mechanism to mitigate spatial inconsistencies caused by asynchronous multimodal feature delivery over cellular networks.

(4) We conducted a comprehensive evaluation of SHARDED using the nuScenes dataset and real-world 5G measurement traces, demonstrating improved trade-offs among perception accuracy, uplink bandwidth consumption, and end-to-end latency.

\vspace*{-1mm}
\section{Brief Background}
\label{sec:bg_problem}

\subsection{Data Fusion Approaches and Algorithms}
\label{sec:why_fusion}

State-of-the-art multi-sensor fusion algorithms are typically designed using one of three fusion approaches: early, intermediate, or late fusion. These three approaches offer distinct trade-offs between information richness, modularity, and robustness. Early fusion first merges raw sensor data before feature extraction, allowing joint representation learning from the outset. While it provides rich cross-modal information, it is susceptible to issues arising from sensor misalignment and modality heterogeneity. 
Examples of early fusion models include ~\cite{mvx_net_multimodal_voxelnet_3d_object} and ~\cite{epnet_enhancing_point_features_image}. Intermediate (feature-level) fusion aims to strike a balance between modularity and performance. It first extracts modality-specific features independently, then merges them into a shared representation space.
Example intermediate fusion models include BEVFusion~\cite{liu2022bevfusion}, CMT~\cite{chen2023deepinteraction}, and 
FocalFormer3D further adopts focal attention to selectively aggregate information from different sensors~\cite{fang2023focalformer3d}.
Late fusion combines final detection outputs from separate modality-specific AI models. This approach is modular and resilient to sensor failure, but often lacks deep semantic integration. Examples of late fusion models include
PointPainting~\cite{pointpainting_sequential_fusion_3d} and DeepFusion~\cite{deepfusion_lidar_camera_multimodal}.

\subsection{5G Network Challenges for AV Teleoperations}
\label{sec:5g_tod_challenges}

\begin{figure}[t]
\vspace{1mm}
\centering

\begin{minipage}[t]{0.46\columnwidth}
    \centering
    \includegraphics[width=\linewidth, height=1.2in]{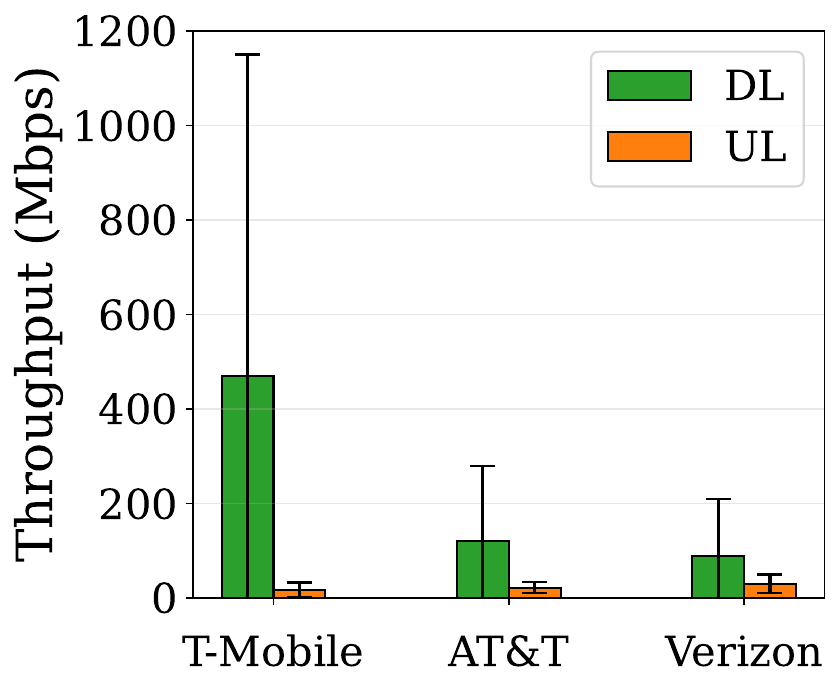}
    \vspace{-1em}
    \caption{Downlink and uplink throughput (Mbps) across major U.S. carriers, showing strong DL--UL asymmetry.}
    \label{fig:5g_asymmetry}
\end{minipage}
\hfill
\begin{minipage}[t]{0.50\columnwidth}
    \centering
    \includegraphics[width=\linewidth, height=1.15in]{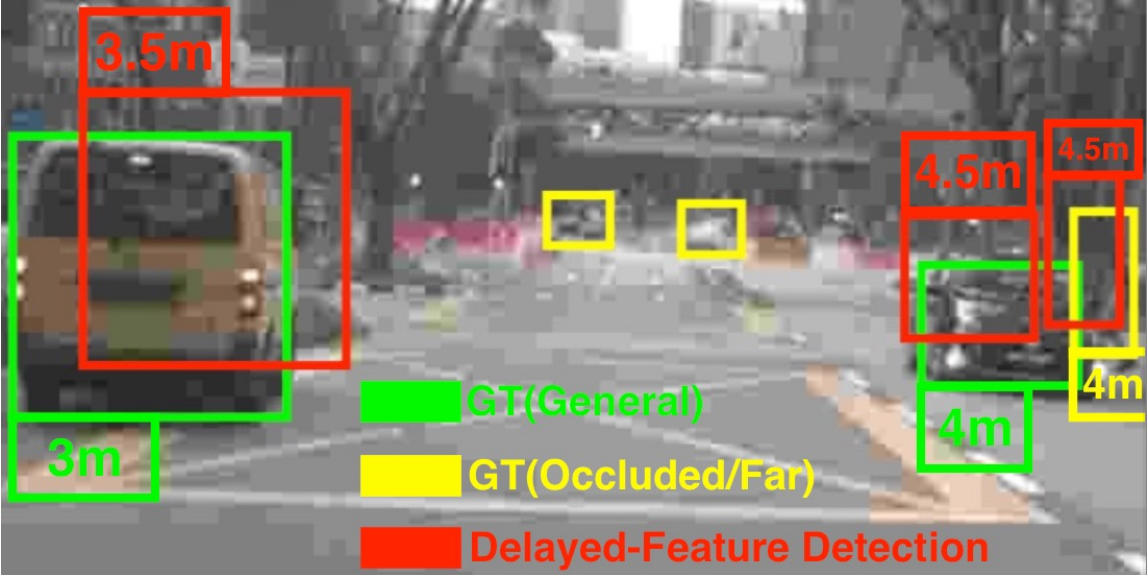}
    \vspace{-1em}
    \caption{Compression-induced perception degradation and latency-induced distance error in remote driving.}
    \label{fig:objs_depth_error}
\end{minipage}

\vspace{-0.8em}
\end{figure}

\noindent
\textbf{Asymmetric Uplink–Downlink Constraints.}
Commercial 5G networks typically prioritize downlink throughput, leaving uplink under-provisioned~\cite{understanding_5g_uplink_constraints, kim2021cellular}.  Fig.~\ref{fig:5g_asymmetry} clearly shows this \emph{downlink vs. uplink asymmetry} using measurement data from three US mobile operators.  This asymmetry is particularly problematic for teleoperation, where uplink traffic dominates: multi-camera video, LiDAR point clouds, and other sensor data streams generate massive amounts of data~\cite{zhang2022uplink,VENTIS23}. Therefore, teleoperation systems face a constant conflict between transmitting rich perceptual information and staying within a feasible uplink budget.

\begin{figure*}[t]
    \centering
    \includegraphics[width=0.95\linewidth]{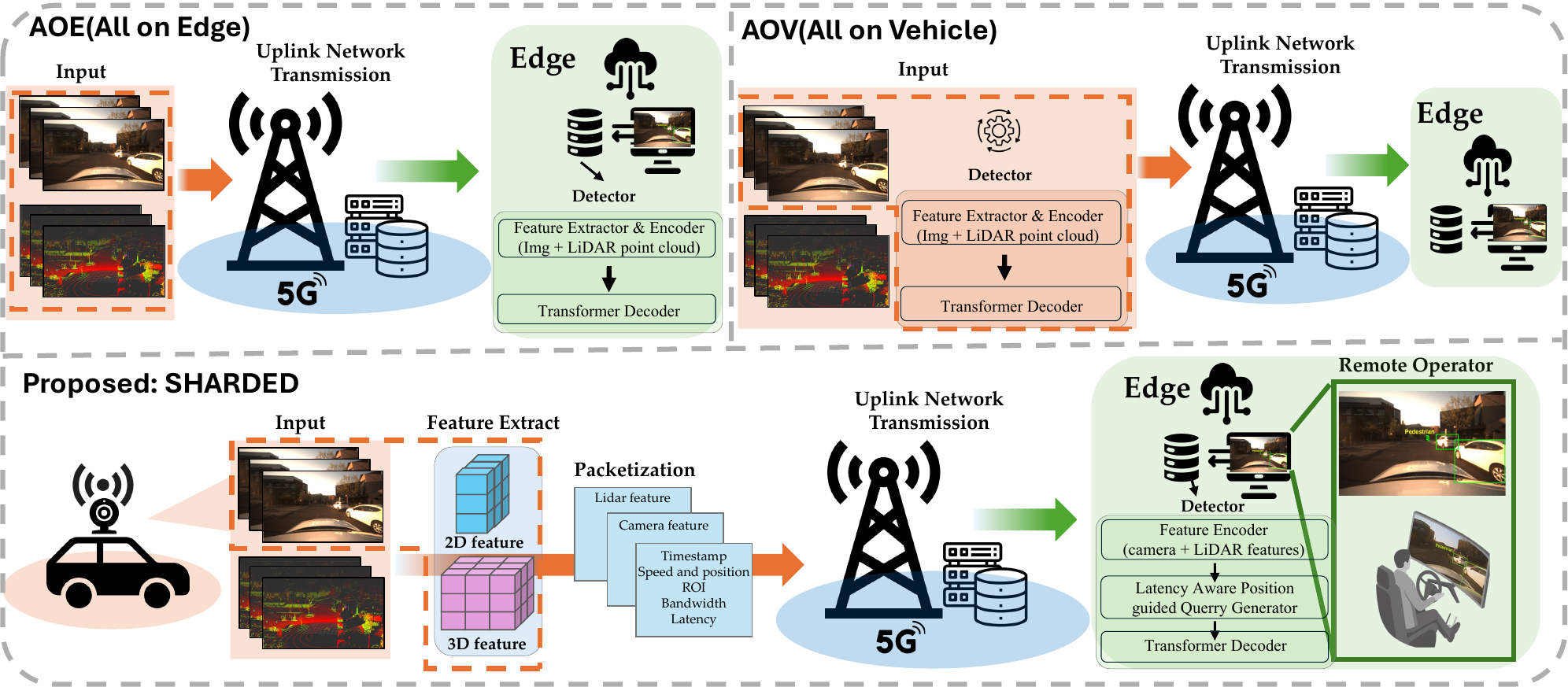}
    \caption{Architecture of the Proposed Approach.}
    
    \label{fig:architecture}
    \vspace{-1.5em}
\end{figure*}

\noindent
\textbf{Network-Induced Perception Challenges.}
Beyond raw bandwidth limitations, teleoperation is also highly sensitive to network-induced latency and jitter.
Throughput fluctuations and transient congestion increase end-to-end latency and disrupt the continuity of remote perception~\cite{throughput_aware_edge_assisted_teleoperation,towards_5g_teleoperated_driving}.
Importantly, heterogeneous sensor data streams often experience different transmission latencies due to differences in data size, encoding processes, and network scheduling. Consequently, the perception module at the edge may receive temporally misaligned inputs; for example, data from one modality (e.g., camera) may arrive earlier than data from another modality (e.g., LiDAR).
This cross-modal asynchrony introduces semantic inconsistencies in remote perception; for instance, delayed depth information can lead to perceived distances to surrounding vehicles that do not match the actual scene.
This problem does not stem from sensor noise or model errors, but rather from latency-induced misalignment under fluctuating 5G network conditions (Fig.~\ref{fig:objs_depth_error}). 
Video bitrate adaptation through compression is the ``go-to'' approach in video streaming to reduce network bandwidth requirements and adapt to its dynamic fluctuation.  Existing research has shown that compressing camera and LiDAR data streams significantly impacts downstream perception results~\cite{Qixin-CQR}, thereby reducing visual clarity and obscuring occluded or long-distance objects for teleoperators (Fig.~\ref{fig:objs_depth_error}). While other methods (see, e.g.,
~\cite{uplink_friendly_teleoperation_with_edge, learning_driven_communication_for_autonomous_driving, intelligent_network_slicing_for_urllc}) have also been proposed in the literature, they are not yet ready to be deployed. Furthermore,  these methods often come at the cost of semantic fidelity.

\begin{table}[t]
\centering
\caption{Comparison of multi-sensor fusion-based perception deployment strategies for teleoperated driving.}
\label{tab:placement_comparison}

\begingroup
\footnotesize
\renewcommand{\arraystretch}{1.25}
\setlength{\tabcolsep}{3pt}

\begin{tabularx}{\columnwidth}{>{\raggedright\arraybackslash}p{0.14\columnwidth}
                              >{\raggedright\arraybackslash}p{0.40\columnwidth}
                              >{\raggedright\arraybackslash}p{0.40\columnwidth}}
\hline
\textbf{Strategy} & \textbf{Advantages} & \textbf{Limitations} \\
\hline
AOV & Low uplink bandwidth demand; independent of networks; simpler design &
Limited onboard compute budget; constrained model capacity \\
AOE & Powerful edge-side compute resources; easy model maintenance and upgrades &
High uplink cost; sensitive to latency and cross-modal misalignment \\
VES & Flexible trade-offs between communication and computation; enables collaborative inference across vehicle and edge & Requires dynamic split design and latency-aware coordination; Need careful balancing of multiple factors and trade-offs; more complex \\
\hline

\end{tabularx}

\vspace{-1.5em}
\endgroup
\end{table}

\vspace*{-1mm}
\section{Multi-Sensor Fusion Deployment Strategies}
\label{sec:problem}
Unlike on-board autonomous driving, where perception, decision-making, and execution all occur in the same location, teleoperated driving (TOD) separates vehicle-side perception from human decision-making at a remote station.
This separation amplifies the impact of perception quality on operator situational awareness and reaction latency.
In complex traffic environments, relying solely on a single perception modality (e.g., only video streams) is often insufficient due to occlusions, depth ambiguity, and unfavorable lighting conditions.
Therefore, multimodal sensor fusion becomes crucial in teleoperated driving, not only to improve perception capabilities but also to enhance robustness and interpretability for human operators.
By integrating complementary modalities such as cameras and LiDAR, fusion techniques enable more reliable spatial reasoning, object localization, and scene understanding. It has been shown that using fused 3D perception results (e.g., object detection and tracking) to augment the operator's view can significantly improve the effectiveness of teleoperated driving.

\subsection{Multi-Sensor Fusion and Perception Deployment Strategies}\label{sec:perception_placement}

The stringent perception demands of teleoperated driving (TOD) and the practical constraints of 5G uplink pose a fundamental design conflict in the multi-sensor, multimodal fusion and perception pipeline design. TOD requires rich, temporally consistent, and semantically reliable perception to support safe remote control. However, 5G uplink limitations restrict both the amount of data that can be transmitted from the vehicle and the timeliness of perceptual updates reaching the operator. As a result, the core challenge is not whether multimodal fusion should be performed, but rather \textit{where the fusion computation should be executed and what representations should be transmitted over the network}.

From a systems perspective, the teleoperation perception pipeline consists of modality-specific feature extraction, cross-modal fusion, and task-level inference.
Model deployment determines how these stages are partitioned between the vehicle and remote computing resources (e.g., edge servers), and which intermediate representations must be transmitted over the network.
Different deployment choices fundamentally impact the trade-offs between semantic fidelity, communication costs, and end-to-end latency.

We first define two baseline deployment strategies for teleoperated perception, namely, All on Vehicle (AOV) and All on Edge (AOE), as shown in Fig.~\ref{fig:architecture}. We then present a third strategy, Vehicle-Edge-Split (VES). 
Table~\ref{tab:placement_comparison} summarizes the advantages and limitations of these deployment strategies.

\noindent \textbf{All on Vehicle (AOV).}
In this strategy, the entire perception and fusion pipeline is executed on the vehicle.
Only lightweight outputs (e.g., detection or tracking results) are transmitted to the remote operator.
AOV minimizes upstream data volume and reduces reliance on network conditions, but places high demands on the vehicle's computing capabilities and limits the complexity of the fusion models that can be deployed on the vehicle.

\noindent \textbf{All on Edge (AOE).}
At the other extreme, perception and fusion computations are primarily performed at the edge.
The vehicle streams raw or minimally processed sensor data upstream, allowing powerful edge-side models to process rich representations.
While AOE benefits from powerful computing capabilities and centralized model management, it generates significant upstream traffic and is highly sensitive to bandwidth fluctuations and network-induced latency.

\noindent \textbf{Vehicle-Edge-Split (VES).}
AOV and AOE represent two extremes of the deployment design space. Between these two extremes there is VES, where the perception pipeline is sharded between the vehicle and the edge.  Partial feature extraction or fusion stages are performed on the vehicle, and intermediate representations are transmitted remotely for further processing. This strategy presents a series of deployment options, allowing the system to trade off uplink bandwidth, computational load, and perception accuracy by adjusting the splitting points in the fusion pipeline. 

Clearly, no single deployment strategy is universally optimal: pipelines that transmit rich intermediate representations can maintain semantic accuracy but increase uplink bandwidth requirements, while more aggressive partitioning strategies can reduce communication costs but may degrade perception performance under poor network conditions.
An appropriate deployment strategy must account for the complex interplay among the fusion structure, network characteristics, and remote-operation latency requirements. In Section~\ref{sec:framework},  we introduce  \emph{SHARDED} -- a flexible and systematic multi-sensor fusion and perception framework designed and optimized for AV teleoperation that enables  \emph{vehicle-edge collaborative model deployment} and allows for \emph{controllable evaluation of various trade-offs}.

\subsection{Related Work}\label{sec:related}
Vehicle-edge compute trade-offs in multi-sensor data fusion and collaborative perception have been studied in the context of connected and autonomous vehicles and vehicle-to-everything (V2X) communications, see, e.g.,~\cite{multi-sensor-edge-computing, multi-modal-data-model-reduction-for-enableing-edge-fusion,qos-content-delivery, occlusion-guided-multimodal-fusion-for-v2i,boosting-collaborative-vehicular-perception,v2x-cooperative-lidar-4dradar-fusion}. The objectives and problem setting of these studies are quite different from ours, as none of them consider AV teleoperation.
Teleoperation introduces additional system-level considerations that fundamentally shape how multimodal fusion should be executed. Fusion of high-dimensional camera and LiDAR data requires substantial computation and data exchange, raising critical questions about where processing should occur and how intermediate representations should be transmitted under bandwidth and latency constraints. In summary, teleoperated driving makes multimodal fusion essential for reliable remote perception, while simultaneously exposing its tight coupling with execution architecture and communication constraints. This dependency becomes particularly critical under practical 5G uplink limitations, which we examine in detail in the following section.

\section{The \textbf{SHARDED} Framework}
\label{sec:framework}
We now present \emph{SHARDED}, a vehicle-edge collaborative camera--LiDAR perception framework for AV teleoperation.

\subsection{SHARDED: Collaborative Vehicle–Edge TOD Perception} 
\label{sec:overall_architecture}
SHARDED structures the perception and sensor streaming pipeline into two parallel data streams: a visual data stream (compressed multi-camera video) for the teleoperator, and a feature-level collaborative perception stream (augmenting the video stream) to support AI-enabled scene understanding.

\subsubsection{Compressed Multi-Camera Visual Data Stream}
Remote teleoperation relies on continuous visual feedback to keep the human operator situationally aware, typically delivered via real-time video streams captured by multiple onboard cameras and transmitted through a video streaming system (e.g., WebRTC, RTSP, or a proprietary protocol).
Due to the uplink bandwidth constraints, each camera video stream must be compressed before transmission. 
Although compressed video enables basic remote control, aggressive compression substantially degrades fine-grained visual details, making it difficult for operators to reliably perceive small, distant, or partially occluded objects. 
To mitigate this limitation, SHARDED provides perception-assisted multimodal support to augment the compressed visual data stream.

Fig.~\ref{fig:architecture} illustrates the overall architecture of SHARDED. 
At the vehicle, camera frames are captured at 15 frames per sec (fps) and streamed via WebRTC with adaptive compression to meet uplink constraints, providing the primary visual input for teleoperation.

\subsubsection{Feature Representation Data Stream for AI-Enabled Edge Inference to Augment Visual Perception}
In parallel with video streaming, SHARDED performs feature-level multimodal perception collaboratively across the vehicle and the edge.
On the vehicle, raw camera images and LiDAR point clouds are processed by modality-specific encoders to extract intermediate feature representations.
The camera encoder produces spatial feature maps that preserve high-level semantics while reducing pixel-level redundancy, whereas the LiDAR encoder converts point clouds into structured feature representations suitable for downstream fusion (e.g., BEV-aligned features).
These encoders correspond to the modality-specific front-ends of a CMT-based fusion backbone. Instead of transmitting raw sensor data, SHARDED transmits the extracted intermediate features to the edge server.
Compared with raw sensor streams, these features retain task-relevant semantic and geometric information while being substantially more compact.
At the edge, the received camera and LiDAR features are processed by the remaining stages of the CMT backbone to perform cross-modal fusion and task-level inference.
Downstream perception heads then generate outputs such as 3D object detection and tracking results. 
The perception results generated at the edge are used to augment the operator’s visual stream as visual overlays, and are not involved in the vehicle control loop.
In particular, SHARDED provides additional depth and object-level cues that are difficult to obtain from compressed video alone, allowing operators to better perceive surrounding objects and spatial relationships and thereby enhancing situational awareness during teleoperation.
This perception assistance remains effective even when the visual stream is degraded by compression or adverse sensing conditions.

Deploying feature-level perception across the vehicle and the edge enables SHARDED to balance communication overhead and computational load.
Vehicle-side feature extraction significantly reduces uplink data volume compared to transmitting raw sensor streams, while executing cross-modal fusion and high-level inference at the edge avoids overloading the vehicle’s on-board compute resources and enables the use of more expressive perception models.
We adopt CMT as the perception backbone because its separation between modality-specific encoding and cross-modal fusion naturally supports such distributed deployment.
Although SHARDED is instantiated with CMT in this work, the proposed framework is applicable to other feature-level camera--LiDAR fusion backbones that separate modality-specific encoding and cross-modal fusion.

\subsection{Adaptive Feature Transmission Under Uplink Constraints}
\label{sec:adaptive_offloading}

While SHARDED performs feature-level deployment, transmitting all intermediate features can still be inefficient under variable network conditions. In teleoperated driving, uplink bandwidth fluctuates, and task-relevant perception is often concentrated near the ego vehicle. This motivates an adaptive feature-transmission mechanism that selectively transmits only task-relevant features, thereby further reducing uplink traffic within a defined budget.

\subsubsection{Distance-Weighted Object Selection}

We formulate adaptive feature transmission at time $t$ as selecting feature regions under an uplink budget:
\begin{equation}
\max_{\mathcal{S}_t}\sum_{i\in\mathcal{S}_t}u_i
\quad \text{s.t.}\quad
\sum_{i\in\mathcal{S}_t} b_i \le B_t,
\label{eq:adaptive_selection}
\end{equation}
where $u_i$ is the utility of region $i$, $b_i$ is its transmission size, $B_t$ is the runtime uplink budget, and $\mathcal{S}_t$ is the set of selected feature regions at time t.

At runtime, SHARDED estimates $B_t$ from recent throughput and latency measurements, then selects the highest-utility LiDAR regions under Eq.~\eqref{eq:adaptive_selection}. The selected LiDAR regions are projected to camera feature space so that aligned camera--LiDAR features are transmitted and fused without modifying downstream perception heads.

To preserve cross-modal spatial consistency for downstream fusion, camera features are selected to correspond to the same spatial regions as the retained LiDAR features. Using calibrated camera--LiDAR extrinsic parameters, the spatial locations of the selected LiDAR features are projected into the camera coordinate system to identify the corresponding regions in the camera feature maps. Only camera feature embeddings aligned with the selected distance band are transmitted.

The selected camera and LiDAR features are transmitted to the edge as part of the perception assistance path. At the edge, the received features are processed by the remaining stages of the fusion backbone and perception heads in the same manner as in the full-feature pipeline. This design keeps the perception model unchanged and treats feature selection purely as a system-level transmission mechanism.

\subsubsection{Runtime Bandwidth-Aware Adjustment}

The distance band serves as an explicit control knob for regulating uplink feature volume under time-varying network conditions. During runtime, SHARDED estimates the available uplink feature budget online based on recent uplink throughput and latency measurements, and adjusts the distance band accordingly. When uplink capacity becomes constrained, the distance range is narrowed to prioritize safety-critical spatial regions; when more bandwidth is available, the range is expanded to improve spatial coverage for teleoperation.

This runtime adjustment exposes an inherent trade-off between communication efficiency and spatial coverage: narrowing the distance band reduces uplink usage but may omit contextual information, whereas expanding the band improves spatial coverage at higher communication cost. By adaptively selecting spatially relevant multimodal features under an explicit uplink budget, SHARDED substantially reduces uplink overhead while preserving task-relevant perception information.

Nevertheless, even with bandwidth-aware feature selection, network latency and its temporal variability remain unavoidable in practical cellular networks, leading to delayed arrival of camera and LiDAR features at the edge. We therefore introduce a complementary latency-aware position drift compensation mechanism to mitigate the resulting cross-modal misalignment.

\subsection{Latency-Aware Position Drift Compensation}
\label{sec:position_drift_cmt}
This section first analyzes the impact of network latency in object position estimation under asynchronous feature delivery. It then presents a refinement strategy that incorporates latency information into the CMT position-guided query generation process, enabling the model to compensate for temporal offsets prior to decoding. Together, these components improve spatial consistency and reduce distance estimation errors under realistic 5G conditions.
\subsubsection{Drift Modeling Under Network-Induced Asynchrony} 
Even with feature-level deployment and adaptive feature transmission, SHARDED remains susceptible to network latency and its temporal variability in commercial 5G networks. Uplink latency is not only non-negligible, but also time-varying due to scheduling dynamics, congestion, and channel conditions. As a result, different sensing modalities may experience different transmission delays, leading to asynchronous arrival of multimodal features at the edge. This cross-modal asynchrony poses a fundamental challenge to reliable multimodal fusion for teleoperated driving.
In the proposed SHARDED approach, compressed camera video is continuously delivered to the operator, while intermediate camera and LiDAR features are transmitted to the edge for perception assistance. Owing to differences in data volume, encoding pipelines, and uplink scheduling, camera and LiDAR features may arrive at the edge at different times. Typically, LiDAR features incur larger and more variable transmission delays under constrained uplink conditions. Consequently, the fusion module may receive camera features reflecting the current scene together with LiDAR features captured at an earlier vehicle pose.
Such temporal misalignment manifests as position drift in fused perception results. Since LiDAR-derived features encode geometric information relative to the vehicle's pose at the time of sensing, delayed LiDAR features may no longer be spatially consistent with newly received camera features. For a moving vehicle, this leads to erroneous estimates of object positions and distances. Importantly, this inconsistency is caused by network-induced delay, rather than sensor noise or model errors, and therefore cannot be addressed by conventional fusion pipelines that assume synchronized inputs.

Let $\tau(t)$ denote the estimated end-to-end transmission latency of the LiDAR features at time $t$. The motion-induced position offset during this interval is approximated using a first-order model as

\vspace{-1.0em}
\begin{equation}
\Delta \mathbf{p}(t) \approx \mathbf{v}\!\left(t-\tau(t)\right)\tau(t),
\label{eq:cmt_drift_offset}
\end{equation}
where $\mathbf{v}(t)=[v_x(t),v_y(t),v_z(t)]^{\top}$ is the ego-vehicle velocity. The compensated position is given by
\begin{equation}
\mathbf{p}_{\mathrm{comp}}(t)
=
\mathbf{p}\!\left(t-\tau(t)\right)
+
\Delta \mathbf{p}(t),
\label{eq:cmt_position_compensation}
\end{equation}
with $\mathbf{p}(t)=[x(t),y(t),z(t)]^{\top}$ denoting the 3D position.

During runtime, SHARDED estimates the effective latency of LiDAR features arriving at the edge based on transmission timestamps and observed network delay. This latency estimate is used to compensate for the position offset caused by ego-vehicle motion during feature transmission.

\subsubsection{Latency-Aware Position Query Refinement}

We incorporate a latency-aware position drift compensation module into the feature-level fusion pipeline to account for relative transmission delay between modalities before cross-modal fusion. At runtime, SHARDED estimates $\tau(t)$ from feature transmission timestamps, computes motion offset using Eq.~\eqref{eq:cmt_drift_offset}, and applies the correction in Eq.~\eqref{eq:cmt_position_compensation} to LiDAR spatial representations before fusing them with current camera features at the edge.

Given the estimated latency, the compensation module adjusts the spatial representation of LiDAR-derived features to align them with the fusion pipeline's current reference frame. This adjustment compensates for vehicle motion during the transmission interval and reduces spatial inconsistency between delayed LiDAR features and timely camera features. In practice, LiDAR-derived spatial features are warped according to the estimated offset $\Delta \mathbf{p}(t)$ in Eq.~\eqref{eq:cmt_position_compensation}, and the compensation module adjusts the 3D spatial coordinates of LiDAR-derived features using the standard CMT fusion module without modifying downstream perception heads.

The latency-aware compensation module is integrated at the edge after feature transmission and before cross-modal fusion. It operates within the perception-assistance path and requires no modifications to the vehicle-side processing pipeline or the adaptive feature transmission mechanism. Because the compensation is applied only to LiDAR spatial representations, it can be adjusted dynamically according to network conditions without affecting other components. By accounting for latency-induced asynchrony prior to fusion, this mechanism improves the temporal consistency of multimodal perception under fluctuating 5G conditions and strengthens the reliability of perception support for teleoperated driving.

\begin{figure*}[t]
\vspace{1mm}
\centering
\begin{minipage}[t]{0.24\linewidth}
    \centering
    \includegraphics[width=\textwidth, height=1.25in]{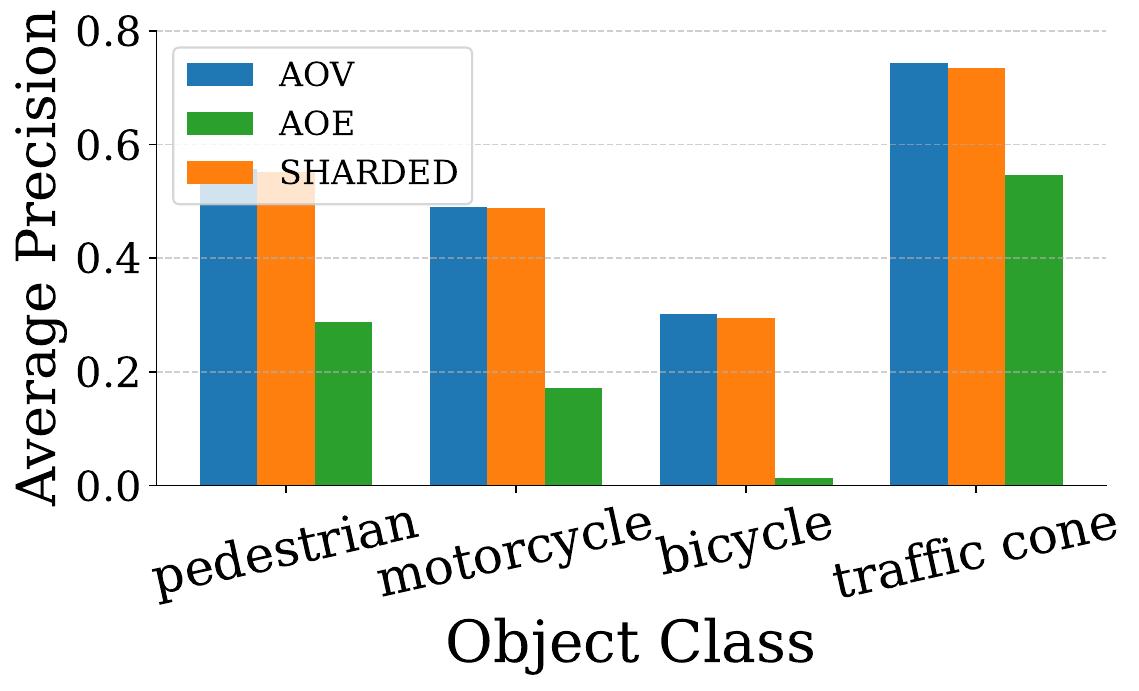}
    \vspace{-2em}
    \caption{Per-category detection accuracy under different deployment strategies.
    }
    \label{fig:obj_classes}
\end{minipage}
\hfill
\begin{minipage}[t]{0.2375\linewidth}
    \centering
    \includegraphics[width=\textwidth, height=1.25in]{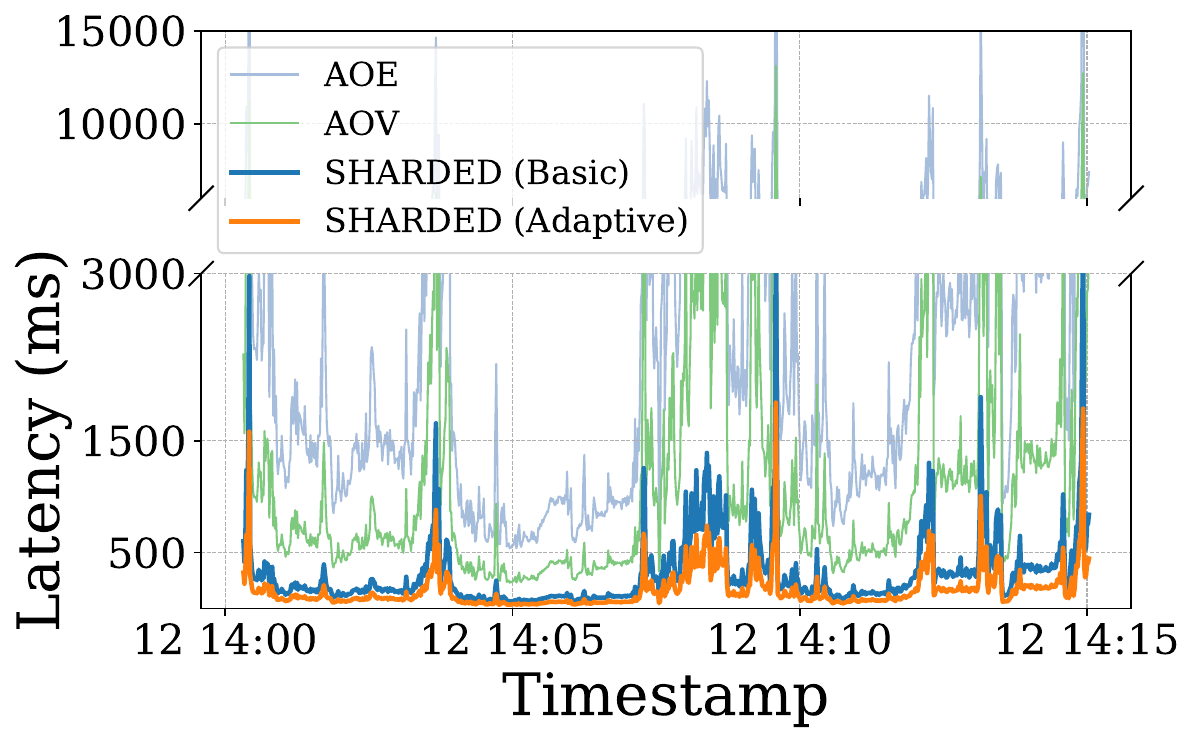}
    \vspace{-2em}
    \caption{End-to-end perception latency of different strategies under real 5G uplink traces.
    }
    \label{fig:latency_over_time}
\end{minipage}
\hfill
\begin{minipage}[t]{0.2375\linewidth}
    \centering
    \includegraphics[width=\textwidth, height=1.25in]{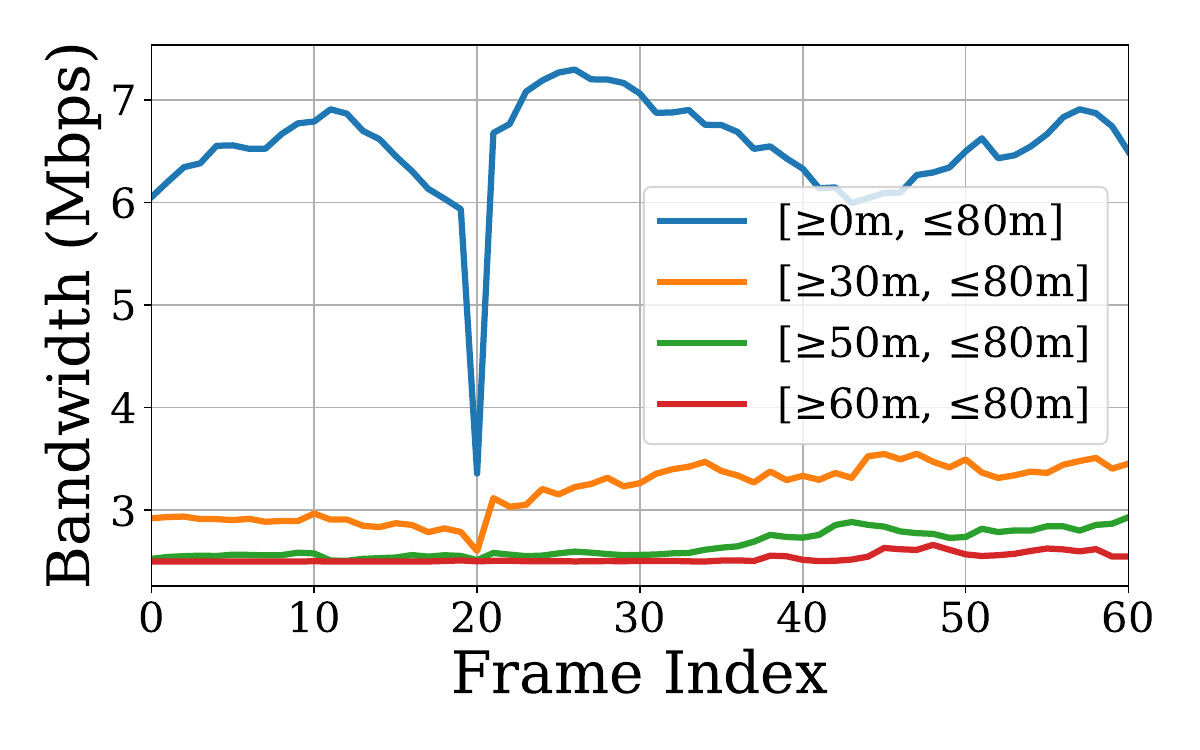}
    \vspace{-2em}
    \caption{Per-frame uplink bandwidth across perception ranges in SHARDED (video streaming excluded).
    }
    \label{fig:frame_bandwidth}
\end{minipage}
\hfill
\begin{minipage}[t]{0.2375\linewidth}
    \centering
    \includegraphics[width=\textwidth, height=1.25in]{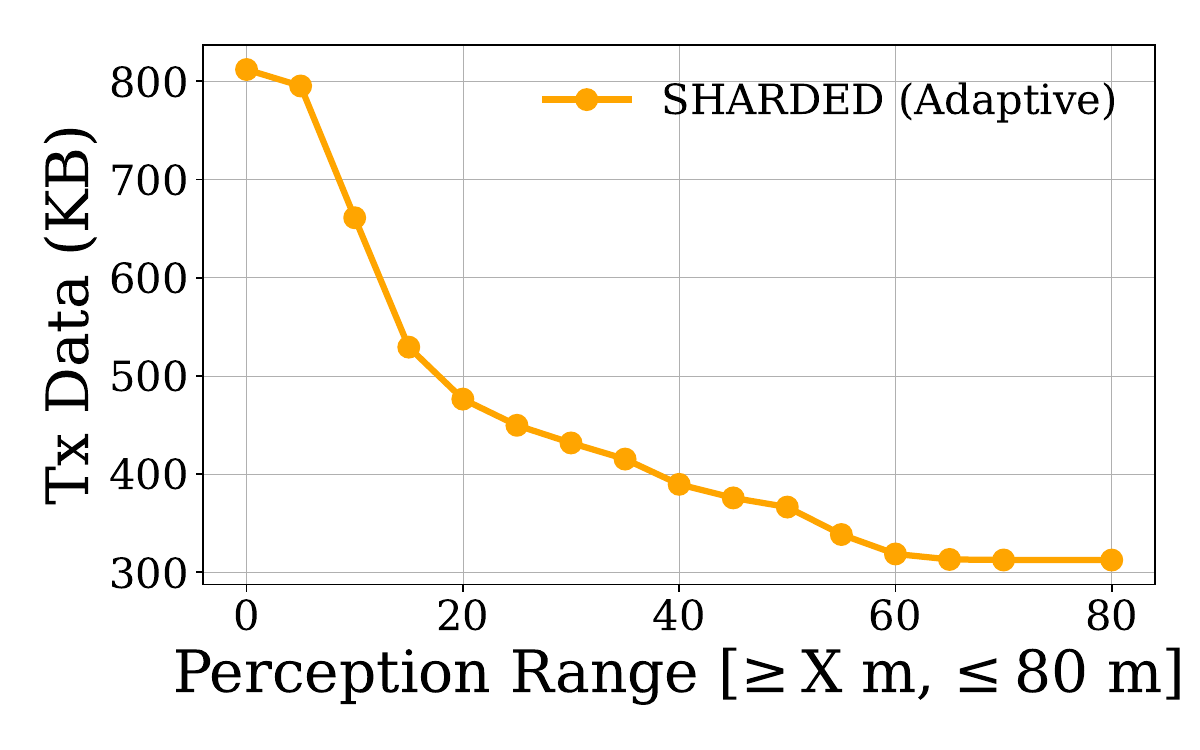}
    \vspace{-2em}
    \caption{Transmitted feature size versus perception range in SHARDED with adaptive transmission mechanism.
    }
    \label{fig:transmitted_data_vs_distance}
\end{minipage}
\vspace{-1.8em}
\end{figure*}

\begin{figure}[t]
\vspace{1mm}
\centering

\begin{minipage}[t]{0.46\columnwidth}
    \centering
    \includegraphics[width=\linewidth, height=1.15in]{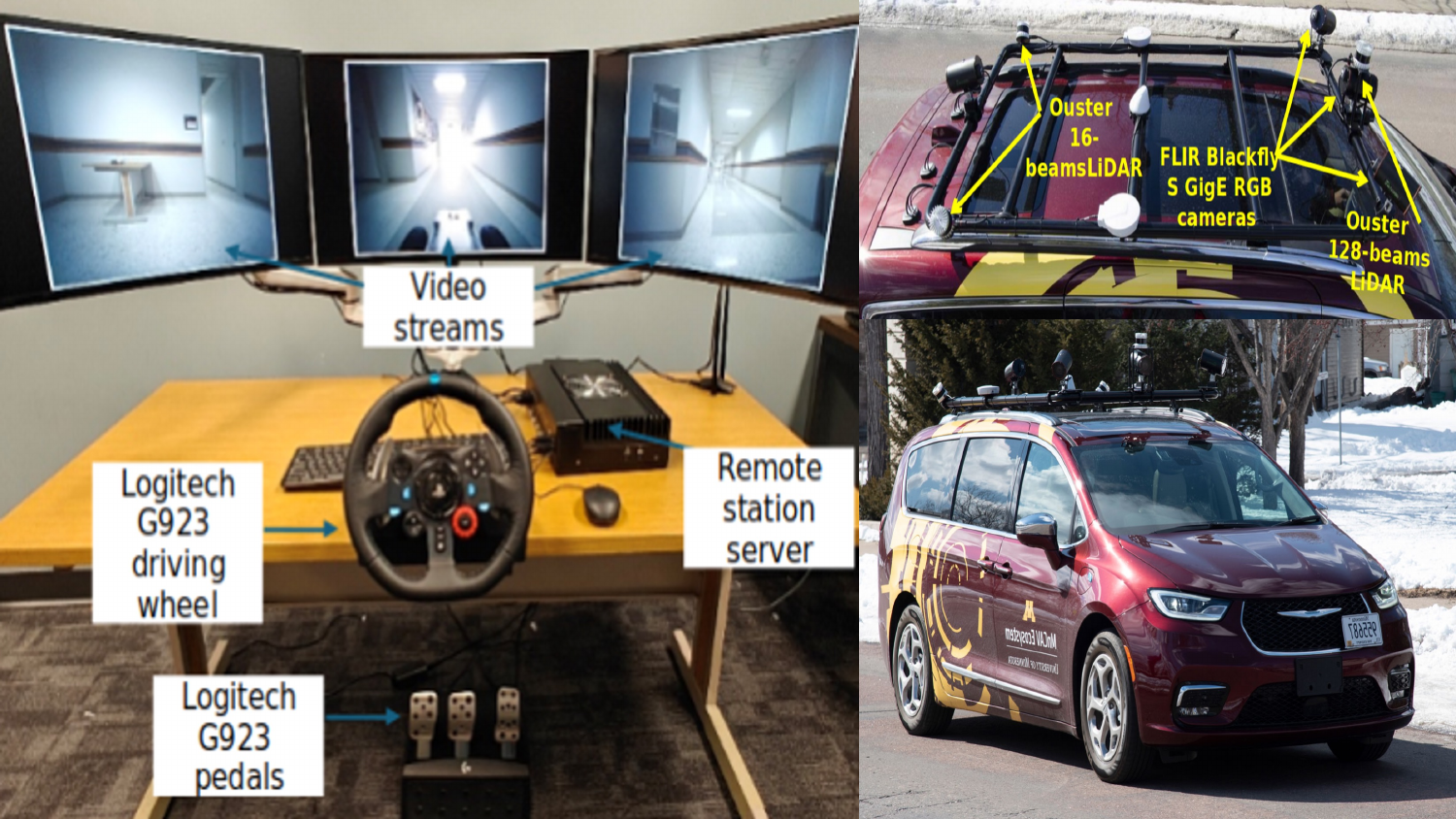}
    \vspace{-1.5em}
    \caption{Real-world teleoperation experimental setup.}
    \label{fig:mncav_vehicle}
\end{minipage}
\hfill
\begin{minipage}[t]{0.50\columnwidth}
    \centering
    \includegraphics[width=\linewidth, height=1.15in]{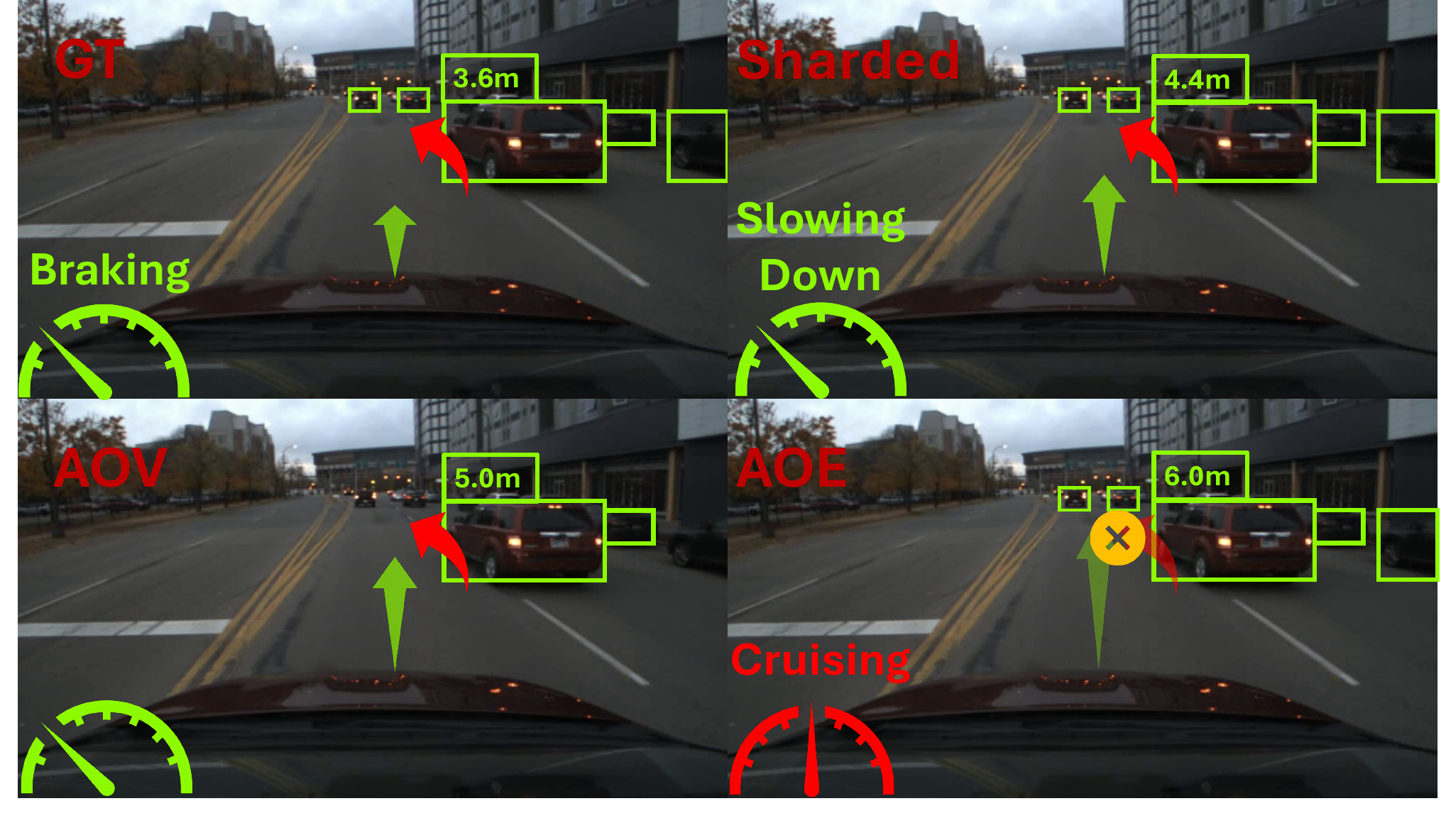}
    \vspace{-1.5em}
    \caption{Visual comparison of baslines and our proposed approach in real world.}
    \label{fig:appraoch_comparison}
\end{minipage}
\end{figure}

\renewcommand{\arraystretch}{1.2}
\setlength{\textfloatsep}{5pt} 
\begin{table}[t]
\centering
\caption{Comparison of AV-to-edge strategies. AOV and AOE transmit a higher-resolution image for operator viewing, while SHARDED transmits only a compressed image, and latency is computed for a perception range of 30-50 meters.}
\begin{tabularx}{\columnwidth}{@{}l c >{\centering\arraybackslash}X >{\centering\arraybackslash}X >{\centering\arraybackslash}X@{}}
\toprule
\textbf{Method} & \textbf{Trans. Data} & \textbf{Transmission Time (ms)} & \textbf{Proc. (ms)} & \textbf{mAP} \\
\midrule
AOV & Results+Img & 321.9 & 215.99 & 0.587 \\
AOE & Sensor Data & 3446.8 & 76.338 & 0.589 \\
\midrule
\makecell[l]{SHARDED\\(Basic)} & Feat+Img & 392.1 & 76.338 & 0.580 \\
\makecell[l]{SHARDED\\(Adaptive)} & Feat+Img & 171.2 & 76.338 & 0.570 \\
\bottomrule
\end{tabularx}
\label{tab:placement_summary}
\vspace{-1ex}
\end{table}

\section{Experimental Evaluation}
\label{sec:evaluation}

\subsection{Experimental Setup}
\label{sec:exp_setup}
We conducted experiments using a MNCAV test vehicle equipped with a 64-channel Ouster LiDAR and six FLIR Blackfly S cameras; the real-world setup is shown in Fig.~\ref{fig:mncav_vehicle}. Data collection included both perception measurements and 5G network performance metrics across diverse driving scenarios, encompassing urban traffic environments and highway conditions. All edge-side perception and fusion computations were performed on a server with an NVIDIA RTX A6000 GPU, and the vehicle-side modules ran on a commercial-grade onboard computing platform (ADLINK ADM-AL30 with NVIDIA RTX 5000 SFF Ada). This setup allowed us to replicate realistic teleoperation constraints, including onboard computational limits and time-varying network conditions, while systematically capturing sensor and network data for subsequent analysis.

\noindent\textbf{Datasets and Methodology.} 
We evaluate SHARDED through trace-driven experiments and real-world deployments, focusing on the trade-offs among perception accuracy, uplink communication overhead, and end-to-end latency in teleoperated driving.
We use the nuScenes dataset~\cite{nuscenes_dataset}, which provides synchronized multi-camera images and LiDAR point clouds with 3D annotations, and evaluate several thousand frames across representative urban scenes and deployment configurations.
We further conduct real-world experiments on a robotic vehicle platform to validate system behavior under realistic sensing noise and network dynamics.

Our evaluation targets system-level deployment and communication trade-offs rather than hardware-specific inference performance.
We emulate commercial cellular uplink conditions by replaying real 5G measurement traces collected from operational networks, comprising over 30 hours of measurements and more than 100,000 throughput and latency samples.
Unless otherwise stated, the operator-facing compressed video stream is always enabled, and all reported uplink bandwidth and transmission costs correspond only to the additional traffic introduced by perception assistance.

\noindent\textbf{Baselines and Metrics.}
We compare SHARDED with two baseline deployment strategies defined in Section~\ref{sec:perception_placement}: all-on-vehicle (AOV) and all-on-edge (AOE).
All deployment modes use the same perception backbone and detection head, and differ only in pipeline placement and the information transmitted over the uplink.
Specifically, AOE transmits compressed perception inputs to the edge for remote fusion, whereas AOV executes the entire pipeline on the vehicle and transmits only final perception results.

We evaluate both perception performance and system efficiency.
Perception accuracy is measured using mean average precision (mAP) over pedestrians, motorcycles, bicycles, and traffic cones.
We additionally report longitudinal distance estimation error along the ego-vehicle.
Communication overhead is quantified by the uplink data volume per frame and the corresponding bandwidth.
We further report end-to-end perception latency from sensor capture to the availability of results at the edge, and decompose it into perception processing latency and feature fusion latency.
Together, these metrics characterize the trade-offs among perception accuracy, responsiveness, and uplink efficiency across deployment modes.

\subsection{Network Performance under Different Deployment Modes}
\label{sec:network_performance}

We conduct an evaluation of the impact of different perception deployment strategies on system performance under realistic uplink constraints in teleoperated driving. Where all three modes (AOE, AOV and SHARDED) employ the same
perception backbone and detection head, and differ only in
pipeline placement and the form of information transmitted
over the uplink. Table~\ref{tab:placement_summary} summarizes their system-level
characteristics in terms of transmitted data volume, processing
cost, end-to-end latency, and perception accuracy. Among three modes, SHARDED adopts a collaborative deployment in which modality-specific feature extraction is performed on the vehicle, while cross-modal fusion and perception heads are executed at the edge, enabling compact intermediate feature transmission and achieving perception performance close to AOV with significantly reduced uplink traffic.


\noindent \textbf{Latency Characteristics and Temporal Stability.}
Fig.~\ref{fig:latency_over_time} illustrates the end-to-end perception latency of the three deployment modes over time. AOE suffers from both higher average latency and more pronounced temporal variability, primarily because large perception inputs must be continuously transmitted over the uplink. By contrast, AOV eliminates uplink transmission of perception inputs and thus achieves lower and more stable latency, albeit at the expense of increased on-board computation. SHARDED attains latency comparable to AOV while preserving perception accuracy close to AOV, thereby providing more timely and temporally stable perception updates for remote operators.

\noindent \textbf{Uplink Bandwidth and Controllability.}
Fig.~\ref{fig:frame_bandwidth} and Fig.~\ref{fig:transmitted_data_vs_distance} jointly show how SHARDED controls uplink cost through perception-range selection. In Fig.~\ref{fig:frame_bandwidth}, each curve corresponds to a range $[X,80]$\,m ($X\in[0,80]$), and the per-frame uplink bandwidth (excluding operator video) increases as the selected range becomes wider (smaller $X$). Fig.~\ref{fig:transmitted_data_vs_distance} explains this trend: wider ranges include more spatial feature regions, so more camera--LiDAR features are transmitted; as $X$ increases, transmitted feature volume decreases monotonically. This coupling between perception range and feature volume enables prioritization of task-relevant regions under uplink constraints, motivating the adaptive feature transmission mechanism in Section~\ref{sec:adaptive_offloading}.

\subsection{Perception Performance and Operator-Centric Metrics}
\label{sec:perception_metrics}


We study how different deployment modes affect perception quality and operator-facing perception consistency in teleoperated driving. 
Unlike fully autonomous driving, teleoperation places a human operator in the control loop, making both perception accuracy and the timeliness of perception updates critical for safe operation. 
Perception errors may prevent operators from noticing small or partially occluded hazards, while temporal inconsistencies in perception can degrade situational awareness.

\noindent \textbf{Perception Accuracy across Object Categories.}
Fig.~\ref{fig:obj_classes} reports the average precision for multiple object categories. 
AOV achieves the highest accuracy since perception is performed directly on locally available sensor data without transmission-induced degradation. 
AOE exhibits noticeably lower accuracy, as the perception inputs must be compressed before uplink transmission, which degrades both geometric and semantic cues. 
SHARDED consistently achieves accuracy close to AOV and substantially higher than AOE, indicating that intermediate features preserve most task-relevant information while avoiding the fidelity loss caused by sensor-stream compression. 
The improvement is particularly important for small and visually inconspicuous objects, such as traffic cones and bicycles, which are difficult for operators to reliably identify from compressed video alone.

We next examine how network-induced latency and cross-modal asynchrony affect operator-facing perception quality, and how the proposed latency-aware position drift compensation improves robustness under delayed feature delivery.

\begin{figure}[t]
    \centering
    \begin{subfigure}[c]{0.68\columnwidth}
        \centering
        \includegraphics[width=\linewidth]{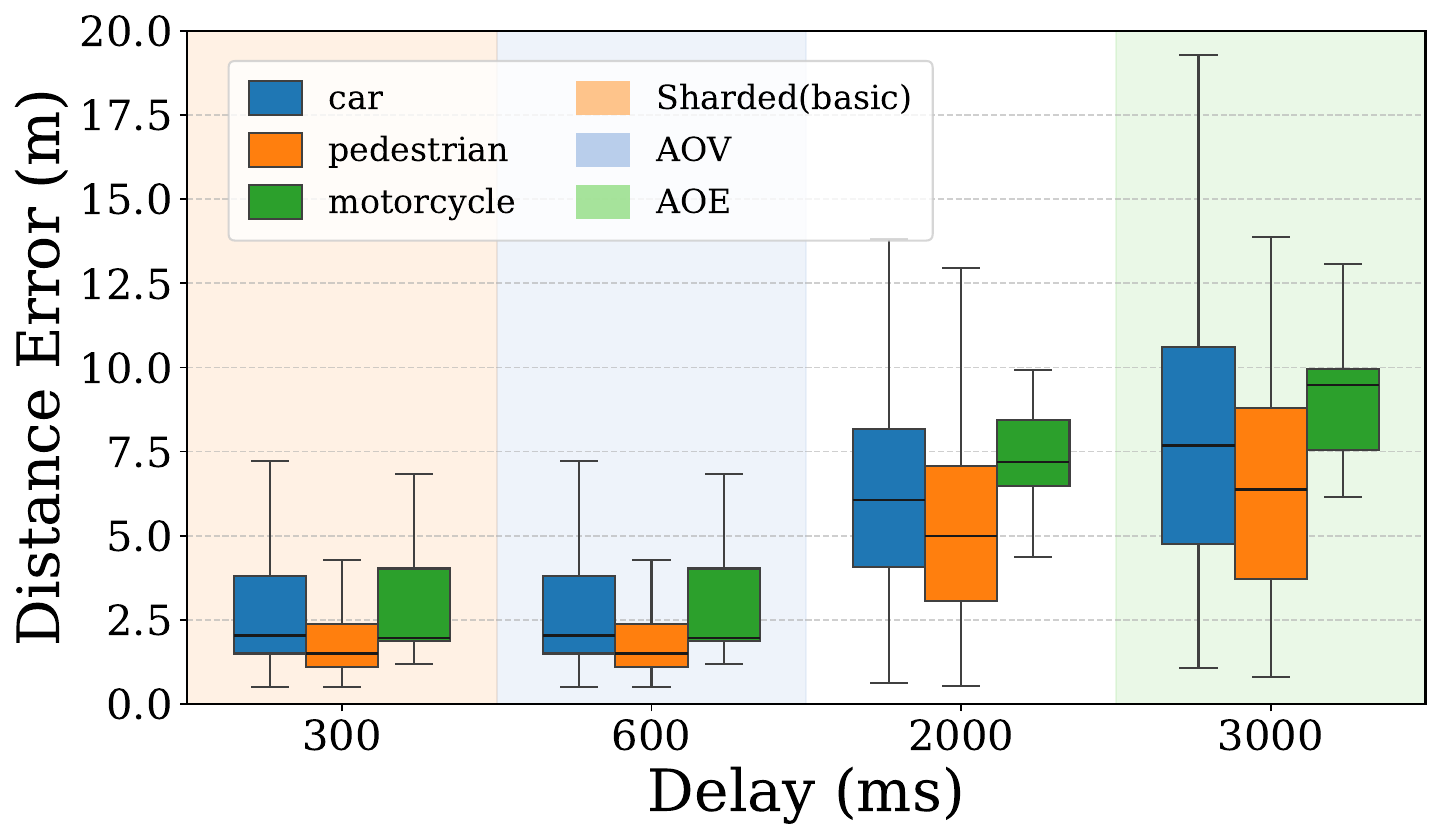}
        \vspace{-1.8em}
        \caption{}
        \label{fig:distance_delay}
    \end{subfigure}
    \hfill
    \begin{subfigure}[c]{0.30\columnwidth}
        \centering
        \includegraphics[width=\linewidth]{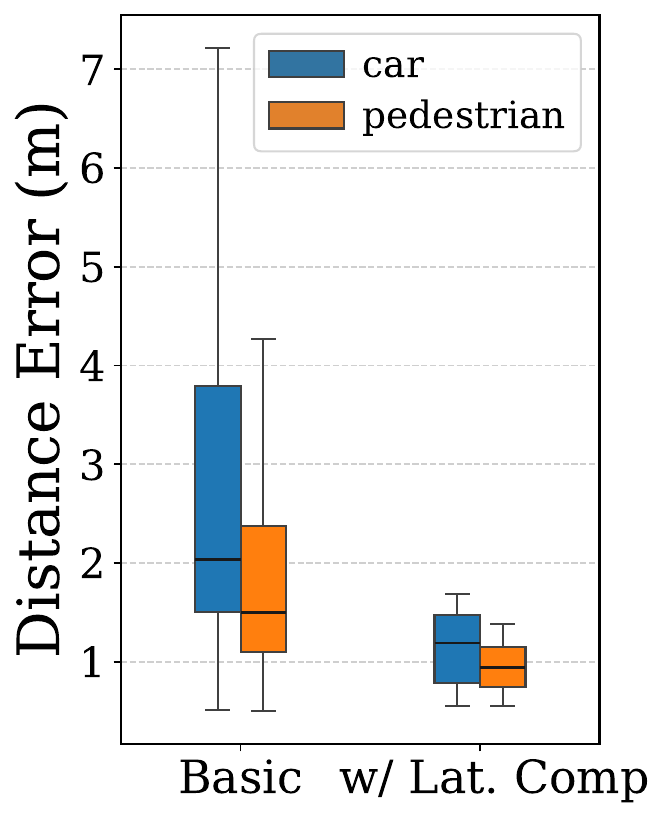}
        \vspace{-1.5em}
        \caption{}
        \label{fig:distance_poscomp}
    \end{subfigure}
    \vspace{-0.5em}
    \caption{Latency-induced longitudinal distance error (left) and the effect of latency compensation in SHARDED (right).}
    \label{fig:distance_error}
    \vspace{-0.6em}
\end{figure}

\noindent \textbf{Quantification of the Impact of Latency on Distance.}
Fig.~\ref{fig:distance_error} illustrates the longitudinal distance estimation error under different network latency levels. 
As uplink latency increases, temporal misalignment between camera and LiDAR features becomes more pronounced, leading to larger distance errors in cross-modal fusion. 
Fig.~\ref{fig:distance_delay} further breaks down the distance estimation error by object category (cars, pedestrians, and motorcycles). Across object categories and latency settings, SHARDED reduces the median longitudinal distance error by up to about 65\% compared with AOE.

Without position drift compensation, SHARDED already exhibits lower distance error than AOV and AOE, since feature-level collaborative perception reduces the amount of transmitted data and shortens the effective perception pipeline. 
However, distance error still increases with latency due to residual cross-modal asynchrony.

\noindent \textbf{Effect of Latency-Aware Position Drift Compensation.}
In teleoperated driving, longitudinal distance estimation directly affects an operator’s ability to judge braking and collision risk when interacting with surrounding objects.
We therefore use longitudinal distance error as a proxy metric to characterize operator-facing depth consistency under network-induced latency. As shown in Fig.~\ref{fig:distance_poscomp}, SHARDED with latency-aware compensation reduces the median longitudinal distance error by approximately 35\%--45\% compared with basic SHARDED for both cars and pedestrians.
This improvement indicates that explicitly compensating for motion-induced spatial offsets during feature transmission effectively mitigates latency-induced geometric inconsistency in multimodal fusion.

Overall, SHARDED achieves higher perception quality under lower uplink bandwidth and latency than both AOV and AOE. Adaptive feature transmission further reduces bandwidth usage and perception latency without degrading detection accuracy, while latency-aware drift compensation mitigates distance estimation errors caused by transmission asynchrony. Together, these mechanisms enable more timely and reliable delivery of task-relevant perception and more consistent depth cues for teleoperation, as illustrated in Fig.~\ref{fig:appraoch_comparison}.

\noindent \textbf{Ablation of SHARDED Components.}
To isolate the contribution of each SHARDED component, we summarize an ablation using the same evaluation setup and existing results. We compare the basic SHARDED pipeline, SHARDED with adaptive transmission, and SHARDED with latency-aware compensation. Adaptive transmission mainly reduces communication cost with negligible impact on mAP, while latency-aware compensation primarily improves distance consistency (a substantial reduction (35\%--45\%) in median longitudinal distance error, normalized to SHARDED Basic; see Fig.~\ref{fig:distance_poscomp}).

\begin{table}[t]
\centering
\caption{Ablation study of SHARDED components.}
\label{tab:ablation_sharded}
\begin{tabularx}{\columnwidth}{@{}l c c c@{}}
\toprule
\textbf{Variant} & \textbf{mAP} & \textbf{Tx Time (ms)} & \textbf{Dist. Err. (m)} \\
\midrule
SHARDED (Basic)                & 0.580        & 392.1 & 1.50--2.00 \\
\makecell[l]{SHARDED + Adaptive Tx}          & 0.570 & 171.2 & 1.50 \\
\makecell[l]{SHARDED + Adaptive Tx \\ \hspace{5.1em}+ Lat. Comp.} & 0.578 & 171.2 & 0.90--1.30 \\
\bottomrule
\end{tabularx}

\vspace{-1mm}
\end{table}

\section{Conclusion}

This paper presents \textbf{SHARDED}, a collaborative camera--LiDAR perception framework for teleoperated driving under realistic 5G uplink constraints. SHARDED deploys a feature-level perception pipeline across the vehicle and the edge, enabling 3D detection and depth-aware perception with substantially reduced uplink communication overhead. Building on this deployment approach, SHARDED further integrates an adaptive feature transmission and latency-aware mechanism to reduce bandwidth demand without degrading detection accuracy, and a latency drift compensation mechanism to mitigate distance errors caused by cross-modal transmission asynchrony. Experimental results using real-world 5G traces show that SHARDED achieves a better balance among detection and depth-aware perception quality, uplink bandwidth consumption, and end-to-end latency than the two baseline deployment strategies, AOV and AOE, demonstrating the effectiveness of feature-level collaborative perception for practical teleoperated driving. 

While SHARDED approach demonstrates strong performance, it relies on range and object type–based feature selection and does not fully capture highly dynamic network fluctuations. Future work will explore context, semantic, and network-aware learning-based adaptation, integrate operator intent, incorporate human-centric studies, and validate at scale in real-world deployments.

\section*{Acknowledgment}
The research was supported in part by NSF awards CNS-2220286, CNS-2220292, CNS-2321531, CNS-2323174, DMS-2436333, and ITE-2453815.

\bibliographystyle{IEEEtran}
\bibliography{bib/refs.bib}

@article{kim2021cellular,
  title={Cellular network support for connected and automated vehicles},
  author={Kim, Hyunwoo and others},
  journal={IEEE Communications Magazine},
  year={2021}
}

@inproceedings{zhang2022uplink,
  title={Characterizing Uplink Performance in 5G Vehicular Networks},
  author={Zhang, Zihan and Feng, Longfei and Wang, Xin and others},
  booktitle = {Proc. IEEE INFOCOM},
  year={2022},
  organization={IEEE}
}

@inproceedings{chen2022cmt,
  title={Cross-Modal Transformer for 3D Object Detection},
  author={Chen, Zetong and others},
  booktitle={CVPR},
  year={2022}
}

@inproceedings{liu2022bevfusion,
  title={BEVFusion: Multi-Task Multi-Sensor Fusion with Unified Bird's-Eye View Representation},
  author={Liu, Zhijian and Tang, Haotian and others},
  booktitle={NeurIPS},
  year={2022}
}

@inproceedings{chen2023deepinteraction,
  title={DeepInteraction: Learning Cross-Modality Interaction for 3D Object Detection},
  author={Chen, Wei and Huang, Bo and Li, Xianglong and others},
  booktitle={CVPR},
  year={2023}
}

@inproceedings{fang2023focalformer3d,
  title={FocalFormer3D: Learning Fine-Grained and Dynamic Representations via Focal Modality Attention for 3D Object Detection},
  author={Fang, Haozhe and Liu, Yu and Zhou, Yuxuan and others},
  booktitle={CVPR},
  year={2023}
}

@inproceedings{mvx_net_multimodal_voxelnet_3d_object,
  title={MVX-Net: Multimodal VoxelNet for 3D Object Detection},
  author={Sindagi, Vishwanath and Zhou, Yin and Tuzel, Oncel},
  booktitle={ICRA},
  year={2019}
}

@inproceedings{epnet_enhancing_point_features_image,
  title={EPNet: Enhancing Point Features with Image Semantics for 3D Object Detection},
  author={Huang, Tao and Gojcic, Zan and others},
  booktitle={ECCV},
  year={2020}
}

@inproceedings{pointpainting_sequential_fusion_3d,
  title={PointPainting: Sequential Fusion for 3D Object Detection},
  author={Vora, Sourabh and Lang, Alex and others},
  booktitle={CVPR},
  year={2020}
}

@inproceedings{deepfusion_lidar_camera_multimodal,
  title={DeepFusion: LiDAR-Camera Deep Fusion for Multi-Modal 3D Object Detection},
  author={Liu, Ziqi and Ma, Xiaoshuai and others},
  booktitle={CVPR},
  year={2023}
}

@inproceedings{understanding_5g_uplink_constraints,
  author    = {Y. Wang and J. Liu and H. Zhao and Z. Qin},
  title     = {Understanding 5G Uplink Constraints for Autonomous Driving},
  booktitle = {IEEE ICC},
  year      = {2022},
  doi       = {10.1109/ICC46239.2022.9838925}
}

@article{learning_driven_communication_for_autonomous_driving,
  author  = {Zhou, Yizhou and Wang, Sen and others},
  title   = {Learning-Driven Communication for Edge-Assisted Autonomous Driving},
  journal = {IEEE Transactions on Mobile Computing},
  year    = {2023},
  volume  = {22},
  number  = {2},
  doi     = {10.1109/TMC.2022.3149819}
}

@inproceedings{throughput_aware_edge_assisted_teleoperation,
  author    = {Kim, Daeyoung and Kim, Seong-Lyun and Ha, Youngju},
  title     = {Throughput-Aware Edge-Assisted Teleoperation for Connected Vehicles},
  booktitle = {IEEE GLOBECOM},
  year      = {2021},
  doi       = {10.1109/GLOBECOM46510.2021.9685782}
}

@inproceedings{towards_5g_teleoperated_driving,
  author    = {Wang, Tao and Yang, Fan and Zhang, Qian and Cao, Jiannong},
  title     = {Towards 5G-Based Teleoperated Driving: System Design and Field Trials},
  booktitle = {IEEE INFOCOM},
  year      = {2020},
  doi       = {10.1109/INFOCOM41043.2020.9155283}
}

@article{intelligent_network_slicing_for_urllc,
  author  = {Chen, Yujiao and others},
  title   = {Intelligent Network Slicing for URLLC in Autonomous Driving},
  journal = {IEEE Transactions on Wireless Communications},
  year    = {2021},
  doi     = {10.1109/TWC.2021.3071373}
}

@inproceedings{uplink_friendly_teleoperation_with_edge,
  author    = {Li, Kai and others},
  title     = {Uplink-Friendly Remote Driving with Edge Collaboration and Semantic Compression},
  booktitle = {ACM MobiCom},
  year      = {2023},
  doi       = {10.1145/3570361.3592524}
}

@misc{5GAA-ToD-system-requirements-analysis,
  title={Tele-Operated Driving (ToD): System Requirements Analysis and Architecture},
  author={5GAA},
  year={2021}
}

@INPROCEEDINGS{Qixin-CQR,
  author={Zhang, Qixin and Sleder, Steven and Hu, Xinyue and others},
  booktitle = {IEEE CQR}, 
  title={Impact of Data Compression on Downstream AI Tasks: A Study using Teleoperated Driving over 5G}, 
  year={2024},
  doi={10.1109/CQR62340.2024.10705883}}

@article{Tangermann2025Waymo-incidents,
  author       = {Victor Tangermann},
  title        = {Elon Musk Scoffs as Rival Waymo Plows Through Car While Driving Wrong Way Down Street},
  journal      = {Futurism},
  year         = {2025},
}

@article{remote-driving-of-autonomous-vehicles-Challenges-and-solutions,
  title={Towards remote driving of autonomous vehicles: Challenges and solutions},
  author={Mahajan, Aishwarya and others},
  journal={IEEE Communications Standards Magazine},
  volume={5},
  number={1},
  year={2021}
}

@inproceedings{Depth-Study-5G-mmWave-Networks,
author="Fezeu, Rostand A. K. and others", title="An In-Depth Measurement Analysis of 5G mmWave PHY Latency and Its Impact on End-to-End Delay", booktitle="PAM", year="2023"
}

@misc{sae2018taxonomy,
  title={{Taxonomy and Definitions for Terms Related to Driving Automation
Systems for On-Road Motor Vehicles, SAE International Recommended
Practice Standard J3016-2018}},
  howpublished={\url{https://www.sae.org/standards/content/j3016_201806/}},
  year={2018}
}

@article{2016-01-0128,
author={Koopman, Philip and Wagner, Michael},
title={{Challenges in Autonomous Vehicle Testing and Validation}},
journal={SAE International Journal of Transportation Safety},
year={2016},
doi={10.4271/2016-01-0128}
}

@misc{how-self-driving-cars-get-help,
    title = {How Self-Driving Cars Get Help From Humans Hundreds of Miles Away},
    note={\url{https://www.nytimes.com/interactive/2024/09/03/technology/zoox-self-driving-cars-remote-control.html}, Last accessed: Sept, 2024}

}

@misc{guident,
  title={Guident Corp. Patented software to improve the safety and availability of autonomous vehicles and robots.},
  author={Guident Corp.},
  year={2024},
  note={\url{https://guident.com/}, Last accessed: Sept 20, 2024}
}

@inproceedings{Ross-SIGCOMM24,
author = {K. Fezeu, Rostand A. and Fiandrino, Claudio and others},
title = {Unveiling the 5G Mid-Band Landscape: From Network Deployment to Performance and Application QoE},
booktitle = {Proc. of ACM SIGCOMM},
year = {2024},
doi = {10.1145/3651890.3672269}
}

@ARTICLE{We-4G-5G-Comparison-journal-paper,
  author={Rochman, Muhammad Iqbal and Ye, Wei and Zhang, Zhi-Li and Ghosh, Monisha},
  journal={IEEE TCCN}, 
  title={A Comprehensive Real-World Evaluation of 5G Improvements Over 4G in Low- and Mid-Bands}, 
  year={2025},
  volume={11},
  number={3},
  doi={10.1109/TCCN.2025.3558062}}

@article{camera-situation-awareness-1,
author={U. {Ramachandran} and K. {Hong} and others}, 
journal={Proceedings of the IEEE}, 
title={Large-Scale Situation Awareness With Camera Networks and Multimodal Sensing}, 
year={2012}, 
doi={10.1109/JPROC.2011.2182093}}

@inproceedings{camera-situation-awareness-2,
  author = {Saurez, Enrique and others},
 title = {Incremental Deployment and Migration of Geo-distributed Situation Awareness Applications in the Fog},
 booktitle = {Proceedings of the 10th ACM International Conference on Distributed and Event-based Systems},
 year = {2016},
 doi = {10.1145/2933267.2933317}
}

@online{TeslaAustinCrash2025,
  author    = {InsideEVs Staff},
  title     = {Tesla Robotaxi Reportedly Crashes Into Parked Car In Austin},
  year      = {2025},
  url       = {https://insideevs.com/news/764905/tesla-robotaxi-first-crash-parked/}
}

@online{WaymoAirportCrash2025,
  author    = {Simon Alvarez},
  title     = {Two Driverless Waymo Cars Collide at Phoenix Sky Harbor Airport},
  year      = {2025},
  url       = {https://www.teslarati.com/two-driverless-waymo-cars-collide-phoenix-sky-harbor-airport/}
}

@inproceedings{VENTIS23,
  author    = {Carpenter, Jason and Ye, Wei and Qian, Feng and
               Zhi{-}Li Zhang},
  title     = {Multi-Modal Vehicle Data Delivery via Commercial {5G} Mobile Networks: An Initial Study},
  booktitle = {VENTIS'23, co-located with IEEE ICDCS'23},
  year      = {2023}
}

@misc{nuscenes_dataset,
      title={nuScenes: A multimodal dataset for autonomous driving}, 
      author={Holger Caesar and others},
      year={2020},
      eprint={1903.11027},
      archivePrefix={arXiv},
      primaryClass={cs.LG},
      url={https://arxiv.org/abs/1903.11027}, 
}

@inproceedings{clocs,
  title={CLOCs: Camera-LiDAR object candidates fusion for 3D object detection},
  author={Pang, Su and others},
  booktitle={2020 IEEE/RSJ International Conference on Intelligent Robots and Systems (IROS)},
  year={2020},
  organization={IEEE}
}

@article{multi-sensor-edge-computing,
  title={Multi-Sensor Data Fusion Meets Edge Computing for Intelligent Surface Vehicles},
  author={Zhang, Boxing and others},
  journal={IEEE Internet of Things Magazine},
  year={2025}
}

@article{multi-modal-data-model-reduction-for-enableing-edge-fusion,
  title={Multi-modal data and model reduction for enabling edge fusion in connected vehicle environments},
  author={Thornton, Samuel and Dey, Sujit},
  journal={IEEE Transactions on Vehicular Technology},
  year={2024},
  publisher={IEEE}
}

@article{qos-content-delivery,
  title={QoS-aware content delivery in 5G-enabled edge computing: Learning-based approaches},
  author={Maleki, Erfan Farhangi and others},
  journal={IEEE Transactions on Mobile Computing},
  year={2024},
  publisher={IEEE}
}

@article{occlusion-guided-multimodal-fusion-for-v2i,
  title={Occlusion-guided multi-modal fusion for vehicle-infrastructure cooperative 3D object detection},
  author={Chu, Huazhen and Liu, Haizhuang and others},
  journal={Pattern Recognition},
  year={2025},
  publisher={Elsevier}
}

@inproceedings{boosting-collaborative-vehicular-perception,
  title={Boosting collaborative vehicular perception on the edge with vehicle-to-vehicle communication},
  author={Zhu, Ruiyang and Zhu, Xiao and others},
  booktitle={Proceedings of the 22nd ACM Conference on Embedded Networked Sensor Systems},
  year={2024}
}

@inproceedings{v2x-cooperative-lidar-4dradar-fusion,
  title={V2X-R: Cooperative LiDAR-4D Radar Fusion with Denoising Diffusion for 3D Object Detection},
  author={Huang, Xun and Wang, Jinlong and others},
  booktitle={Proceedings of the Computer Vision and Pattern Recognition Conference},
  year={2025}
}

\end{document}